\documentclass[10pt,conference, a4paper]{IEEEtran}

\usepackage{cite}
\usepackage{graphicx}
\usepackage{psfrag}
\usepackage{subfigure}
\usepackage{flushend}
\usepackage{color}

\usepackage{amsmath}
\usepackage{array}
\usepackage{graphics}
\usepackage{epsfig}
\usepackage{amsbsy}
\usepackage{amssymb}
\usepackage{amsthm}

\usepackage{framed}
\usepackage{algorithm}
 
\IEEEoverridecommandlockouts

\begin{document}

\title{\LARGE Secondary Outage Analysis of Amplify-and-Forward\\ Cognitive Relays with Direct Link and Primary Interference}
\author{Subhajit~Majhi$^\dagger$,
Sanket S.~Kalamkar$^*$,        
        and~Adrish~Banerjee$^*$
        \vspace*{-4mm}

\thanks{$^\dagger$The author is with the Department of Electrical and Computer Engineering, University of Waterloo, Waterloo, Canada (e-mail: smajhi@uwaterloo.ca).}
\thanks{*The authors are with the Department of Electrical Engineering, IIT Kanpur, India (e-mail: kalamkar@iitk.ac.in, adrish@iitk.ac.in).}
\thanks{Sanket S. Kalamkar is supported by the Tata Consultancy Services (TCS) research fellowship.}

}

\maketitle

\begin{abstract}
The use of cognitive relays is an emerging and promising solution to overcome the problem of spectrum underutilization while achieving the spatial diversity. In this paper, we perform an outage analysis of the secondary system with amplify-and-forward relays in a spectrum sharing scenario, where a secondary transmitter communicates with a secondary destination over a direct link as well as the best relay. Specifically, under the peak power constraint, we derive a closed-form expression of the secondary outage probability provided that the primary outage probability remains below a predefined value. We also take into account the effect of primary interference on the secondary outage performance. Finally, we validate the analysis by simulation results.
\end{abstract}
\vspace*{-2mm}
\section{Introduction}\vspace*{-2mm}

\subsection{Relays in Cognitive Radio}
In future wireless networks, cognitive radio~\cite{mitola} is an exciting solution to overcome the inefficient use of spectrum as it allows spectrum sharing between the licensed user (primary user) and the unlicensed user (secondary user). In a spectrum sharing scenario~\cite{qing,xing}, a secondary user (SU) may share the spectrum with the primary user (PU), provided that SU does not violate the interference constraint at the PU receiver$-$which prompts SU to limit its transmit power to satisfy the interference constraint. 

The use of relays for secondary communication in cognitive radio, at the same time, offers better reliability and improved coverage for SU's transmission~\cite{luo,zhang1,zou,duong,lee}. In addition, the cognitive relays provide increased spatial diversity compared to only direct link transmission. However, the secondary system with relays, in spectrum sharing, faces particularly following two challenges that hinder its performance:
\begin{itemize}
\item[1)] Limitations on its transmit power to satisfy the interference constraint at PU receiver.
\item[2)] Harmful interference from primary transmissions.
\end{itemize}

Among various relaying protocols, amplify-and-forward (AF) and decode-and-forward (DF) are the most popular due to their low complexity. In AF relaying, a relay amplifies the signal received from the secondary  transmitter and forwards it to the secondary destination~\cite{sun,li}, whereas in DF relaying, the relay decodes the received signal and forwards it to the secondary destination~ \cite{khaled,zou}.
\subsection{Contributions and Related Work}\vspace*{-1mm}
1) \textit{Contributions}: We perform an analysis for the outage probability of a secondary system with AF relaying, provided that the outage probability of PU remains below a predefined threshold$-$we characterize the interference to PU as its outage probability. We couple the primary outage constraint with the peak power constraint. We then choose the best relay that maximizes the end-to-end signal-to-interference noise ratio (SINR), and derive a closed-form expression for the secondary outage probability considering the interference from the primary transmission. We assume the presence of the direct link between the secondary transmitter and the secondary destination, and use the maximum ratio combining (MRC) to combine two copies of signal$-$one via direct link and second via the best relay$-$at the secondary destination. 

2) \textit{Related Work}: In~\cite{zou,si11}, authors derive a closed-form expression of the secondary outage probability with the direct link and primary interference under PU's outage probability constraint. In~\cite{duong2}, authors consider a spectrum sharing scenario, where a single AF relay assists the secondary direct link communication, and the signals at the secondary destination are combined by selection combining; but the PU interference is ignored. In~\cite{duong3}, authors study the effect of PU's interference on secondary outage probability for AF relays in absence of the direct link, while~\cite{xu} uses similar setup like~\cite{duong3} for DF relays. Authors in~\cite{yang,huang} study a secondary system with DF relays under direct link and primary interference with the interference power constraint at PU.  The references~\cite{luo,yan} consider the direct secondary link along with DF relays and calculate the secondary outage probability. However, they ignore the effect of PU's interference on the secondary transmission.\vspace*{-2mm}

\section{System Model}

\begin{figure}
\centering
\includegraphics[scale=0.19]{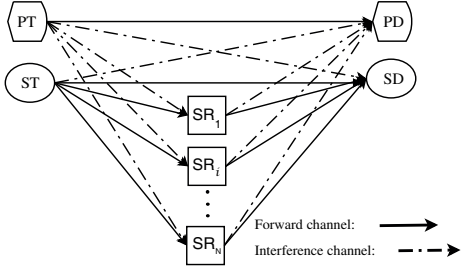}\vspace*{-3mm}
\caption{Secondary transmissions via AF relays in spectrum sharing.}
\label{fig:syst}\vspace*{-5mm}
\end{figure}

Consider a cognitive radio network consisting of a primary transmitter\,(PT), a primary destination\,(PD), a secondary transmitter\,(ST), a secondary destination\,(SD), and $N$ AF secondary relays\,(SR), as shown in Fig.\,\ref{fig:syst}. The ST communicates with SD via the direct link as well as $i$th AF relay ($i =$ 1, 2, $\dotsc$, $N$). The relays operate in a half-duplex mode. The communication between ST and SD happens over two time slots, each of $T$-second duration. In the first time slot, ST transmits the signal with power $P_{\mathrm{ST}}$ to SD over the direct link, and to secondary relays; while in the second time slot, the best relay amplifies the received signal and forwards it to SD with power $P_{\mathrm{SR}_i}$. At SD, two received signal copies$-$first via direct link and second via the best relay$-$are combined by the maximum ratio combining. Relay selection can be employed by a centralized entity, such as the secondary source or a secondary network-manager or in a distributed manner using timers~\cite{shah}. We consider the peak power constraint $P_{\mathrm{pk}}$ on transmit powers of  ST and $i$th secondary relay. In addition, the constraint that the primary outage probability should be below a predefined value regulates the transmit powers of ST and $i$th secondary relay. Denote the powers of ST and $i$th secondary relay, when they are regulated by the primary outage constraint alone, by $P_{\mathrm{u,ST}}$ and $P_{\mathrm{u,SR}_i}$, respectively. Then, combining both above constraints, the maximum allowable powers for ST and $i$th secondary relay become\vspace*{-1mm}
\begin{equation}
P_{\mathrm{ST}} = \min\left(P_{\mathrm{pk}}, P_{\mathrm{u,ST}}\right)
\label{eq:stp}
\end{equation}
and \vspace*{-1mm}
\begin{equation}
P_{\mathrm{SR}_i} = \min\left(P_{\mathrm{pk}}, P_{\mathrm{u,SR}_i}\right),
\label{eq:srp}\vspace*{-1mm}
\end{equation}
respectively. The channel between a transmitter $a \in \lbrace \text{PT}, \text{ST}, \text{SR}_i\rbrace$ and a receiver $b \in \lbrace \text{PD}, \text{SD}, \text{SR}_i\rbrace$ is a Rayleigh fading channel with its channel gain denoted by $h_{a-b}$. Therefore, the channel power gain $|h_{a-b}|^2$ is exponentially distributed with the mean channel power gain $\Omega_{a-b}$. Thus, we can write the probability density function (PDF) and cumulative distribution function (CDF) of $X = |h_{a-b}|^2$ as\vspace*{-2mm}
\begin{equation}
f_{X}(x) =\frac{1}{\Omega_{a-b}}\exp\left(-\frac{x}{\Omega_{a-b}}\right), x \geq 0,
\label{eq:pdf}\vspace*{-1mm}
\end{equation}
\begin{equation}
F_{X}(x) = 1 - \exp\left(-\frac{x}{\Omega_{a-b}}\right), x \geq 0,
\label{eq:cdf}\vspace*{-1mm}
\end{equation}
respectively, where $\exp(\cdot)$ represents the exponential function. We consider that the channels are independent of each other, experience block-fading, and remain constant for two slots of the secondary communication, i.e., for $2T-$second, as in \cite{luo, zou, duong}.\vspace*{-1mm}

\section{Maximum Average Allowable Transmit Power for Secondary Transmitter and Relays}
We use the primary outage probability to characterize the quality of service (QoS) of primary transmissions. The outage probability of the primary user should be below a certain value $\lambda_{\mathrm{p}}$, given the interference from the secondary transmitter and relay. For a constant primary transmit power $P_{\mathrm{PT}}$, we can calculate the primary outage probability as\vspace*{-1mm}
\begin{equation}
\mathrm{P_{out_{p}}}\!\! = \mathrm{Pr}\!\left(\!\log_{2}\left(1+\frac{P_{\mathrm{PT}}|h_{\mathrm{PT-PD}}|^2}{P_{\mathrm{t,ST}}|h_{\mathrm{ST-PD}}|^2 + N_0}\right)\!< R_{\mathrm{p}}\!\right) \leq \lambda_{\mathrm{p}},
\label{eq:3}\vspace*{-1mm}
\end{equation}
where $R_{\mathrm{p}}$ is the primary user's desired data rate, $N_0$ is additive white Gaussian noise (AWGN) power at all receivers, and $P_{\mathrm{t,ST}}$ is the transmit power of ST. The term $\frac{P_{\mathrm{PT}}|h_{\mathrm{PT-PD}}|^2}{P_{\mathrm{t,ST}}|h_{\mathrm{ST-PD}}|^2 + N_0}$ represents the received SINR at PD. In \eqref{eq:3}, at the maximum allowed average power $P_{\mathrm{u,ST}}$ for ST, i.e., when $P_{\mathrm{t,ST}}$ = $P_{\mathrm{u,ST}}$, the weak inequality becomes equality. Thus, from \eqref{eq:3}, conditioned on $|h_{\mathrm{ST-PD}}|^2 = x$, we can write\vspace*{-1mm}
\begin{equation}
\mathrm{P_{out_{p}}}\bigg|_{|h_{\mathrm{ST-PD}}|^2 = x}\hspace*{-5mm} = \mathrm{Pr}\bigg(\!|h_{\mathrm{PT-PD}}|^2 \!<\! \frac{\theta_{\mathrm{p}} (P_{\mathrm{u,ST}}x + N_0)}{P_{\mathrm{PT}}}\!\bigg) = \lambda_{\mathrm{p}},\vspace*{-1mm}
\label{eq:out}
\end{equation}
where $\theta_{\mathrm{p}} = 2^{R_{\mathrm{p}}}-1$. Thus, we can write \eqref{eq:out} as\vspace*{-1mm}
\begin{equation}
\mathrm{P_{out_{p}}}\bigg|_{|h_{\mathrm{ST-PD}}|^2 = x}= 1 - \mathrm{exp} \left( - \frac{\theta_{\mathrm{p}} (P_{\mathrm{u,ST}}x + N_0)}{\Omega_{\mathrm{PT-PD}} P_{\mathrm{PT}}} \right).\vspace*{-2mm}
\label{eq:out1}
\end{equation}
Taking expectation with respect to $|h_{\mathrm{ST-PD}}|^2 $, we obtain\vspace*{-1mm}
\begin{equation}
\mathrm{P_{out_{p}}} = 1 - \frac{\mathrm{exp} \left( - \frac{\theta_{\mathrm{p}}  N_0}{\Omega_{\mathrm{PT-PD}} P_{\mathrm{PT}}} \right)}{1 + \frac{\theta_{\mathrm{p}}  P_{\mathrm{u,ST}} \Omega_{\mathrm{ST-PD}} }{\Omega_{\mathrm{PT-PD}} P_{\mathrm{PT}}}}.\vspace*{-1mm}
\end{equation} 
Rearranging the terms and using \eqref{eq:out}, we find the maximum secondary transmit power $P_{\mathrm{u,ST}}$ under alone primary outage constraint as\vspace*{-2mm}
\begin{equation}
P_{\mathrm{u,ST}} = \frac{P_{\mathrm{PT}} \Omega_{\mathrm{PT-PD}}}{\theta_{\mathrm{p}} \Omega_{\mathrm{ST-PD}}} \left( \frac{\mathrm{exp} \left( \frac{\theta_{\mathrm{p}} N_0}{\Omega_{\mathrm{PT-PD}} P_{\mathrm{PT}}} \right)}{1-\lambda_{\mathrm{p}}} -1 \right)^+, \label{eq:SSpower}\vspace*{-1mm}
\end{equation} 
where $(x)^+ = \max (x,0)$. After combining with the peak power constraint, the maximum average allowable transmit power $P_{\mathrm{ST}}$ for the secondary transmitter can be given by \eqref{eq:stp}. Similar to \eqref{eq:SSpower}, the transmit power of $i$th secondary relay regulated alone by the primary outage constraint can be readily found as\vspace*{-2mm}
\begin{equation}
P_{{\mathrm{u, SR}}_i} = \frac{P_{\mathrm{PT}} \Omega_{\mathrm{PT-PD}}}{\theta_{\mathrm{p}} \Omega_{\mathrm{SR}_i-\mathrm{PD}}} \left( \frac{\mathrm{exp} \left( \frac{\theta_{\mathrm{p}} N_0}{\Omega_{\mathrm{PT-PD}} P_{\mathrm{PT}}} \right)}{1-\lambda_{\mathrm{p}}} -1 \right)^+. \label{eq:SRipower}\vspace*{-1mm}
\end{equation}
After combining with the peak power constraint, the maximum average allowable transmit power $P_{\mathrm{SR}_i}$ for relay $i$ can be given by \eqref{eq:srp}.\vspace*{-1mm}

\section{Derivation of Secondary Outage Probability}\vspace*{-1mm}
The AF relays cooperate opportunistically, where the relay with the largest end-to-end SINR at the secondary destination is selected to forward the received signal in the second time slot. Thus, after receiving the signal from both time slots, SD combines them using MRC technique. The end-to-end SINR is given by \cite{ikki,si}\vspace*{-2mm}
\begin{align}
\gamma_{\mathrm{eq}} &= \gamma_{\mathrm{SD}} + \max_{\mathrm{SR}_i \in \mathbb{R}}\left(\frac{\gamma_{\mathrm{SR}_i}\gamma_{\mathrm{R}_i\mathrm{D}}}{1 + \gamma_{\mathrm{SR}_i}+\gamma_{\mathrm{R}_i\mathrm{D}}}\right) \nonumber \\
&\leq  \gamma_{\mathrm{SD}} + \max_{\mathrm{SR}_i \in \mathbb{R}}(\min(\gamma_{\mathrm{SR}_i}, \gamma_{\mathrm{R}_i\mathrm{D}})) = \gamma_{\mathrm{tot}},
\label{eq:rn1}
\end{align}\vspace*{-5mm}

\noindent where $\mathbb{R}$ is the set of relays given as $\mathbb{R} = \lbrace \mathrm{SR}_1, \dotsc, \mathrm{SR}_i, \dotsc, \mathrm{SR}_N\rbrace$, $\gamma_{\mathrm{SR}_i}$, $\gamma_{\mathrm{SD}}$, and $\gamma_{\mathrm{R}_i\mathrm{D}}$ denote SINR at the $i$th relay, and SINR at SD due to direct transmission and relaying respectively, which are given by\vspace*{-1mm}
\begin{equation}
\gamma_{\mathrm{SR}_i} =  \frac{P_{\mathrm{ST}}|h_{\mathrm{ST-SR}_i}|^2}{P_{\mathrm{PT}}|h_{\mathrm{PT-SR}_i}|^2 + N_0},
\end{equation}\vspace*{-1mm}
\begin{equation}
\gamma_{\mathrm{SD}} =  \frac{P_{\mathrm{ST}}|h_{\mathrm{ST-SD}}|^2}{P_{\mathrm{PT}}|h_{\mathrm{PT-SD}}|^2 + N_0},
\end{equation}\vspace*{-1mm}
\begin{equation}
\gamma_{\mathrm{R}_i\mathrm{D}} =  \frac{P_{\mathrm{SR}_i}|h_{\mathrm{SR}_i-\mathrm{SD}}|^2}{P_{\mathrm{PT}}|h_{\mathrm{PT-SD}}|^2 + N_0}.
\end{equation}
For analytical tractability, we use the upper bound given in \eqref{eq:rn1}, which is tight in medium to high SINR range \cite{ikki,si}. We can obtain $P_{\mathrm{ST}}$ and $P_{\mathrm{SR}_i}$ from \eqref{eq:stp} and \eqref{eq:srp}. The secondary outage occurs when the instantaneous SINR of the secondary transmission falls below the designated threshold, $\theta_\mathrm{s}$. Thus, we can write the secondary outage probability as\vspace*{-4mm}

\begin{eqnarray}
\mathcal{P}_o &=& \mathrm{Pr}(\gamma_{\mathrm{SD}} + \max_{\mathrm{SR}_i \in \mathbb{R}}(\min(\gamma_{\mathrm{SR}_i}, \gamma_{\mathrm{R}_i\mathrm{D}})) < \theta_{\mathrm{S}}),
\end{eqnarray}\vspace*{-4mm}

\noindent where $\theta_{\mathrm{S}} = 2^{2{R}_\mathrm{S}} - 1$ with ${R}_\mathrm{S}$ is the desired secondary data rate.
From \eqref{eq:rn1}, we can see that, $\gamma_{\mathrm{SD}}$, $\gamma_{\mathrm{R}_i\mathrm{D}}$, and $\gamma_{\mathrm{R}_j\mathrm{D}}$ ($i \neq j$) contain a common term $|h_{\mathrm{PT-SD}}|^2$, that makes them dependent. Thus, conditioning on $|h_{\mathrm{PT-SD}}|^2 = y$ and denoting $Z = \underset{\mathrm{SR}_i \in \mathbb{R}}{\max}(\min(\gamma_{\mathrm{SR}_i}, \gamma_{\mathrm{R}_i\mathrm{D}})) $, we can write\vspace*{-2mm}
\begin{eqnarray}
\mathrm{Pr}(\gamma_{\mathrm{tot}} < \theta_{\mathrm{S}}) \big|_{|h_{\mathrm{PT-SD}}|^2 = y} \!\!\!\!\!&=&\!\!\!\!\! \mathrm{Pr} \left( \gamma_{\mathrm{SD}} < \theta_{\mathrm{S}} - Z \right)\nonumber \\
\!\!\!\!\!&=&\!\!\!\!\! \int_{0}^{\theta_{\mathrm{S}}} \!\!\!F_{\gamma_{\mathrm{SD}}} \left( \theta_{\mathrm{S}} - z \right) f_Z(z) \mathrm{d}z.\nonumber\\
\label{eq:main}
\end{eqnarray}\vspace*{-8mm}

\noindent Now, we have
\begin{eqnarray}
F_{\gamma_{\mathrm{SD}}}(z)\big|_{|h_{\mathrm{PT-SD}}|^2 = y} \!\!\!\!&=&\!\!\!\! \mathrm{Pr}\left( |h_{\mathrm{ST-SD}}|^2 < \frac{z(P_{\mathrm{PT}}y + N_0)}{P_{\mathrm{ST}}}\right) \nonumber \\
\!\!\!\!&=&\!\!\!\! 1 - \exp \left( -\frac{z(P_{\mathrm{PT}}y + N_0)}{\Omega_{\mathrm{ST-SD}} P_{\mathrm{ST}}} \right).
\end{eqnarray}\vspace*{-5mm}

\noindent We also have\vspace*{-1mm}
\begin{eqnarray}
 F_Z(z)\big|_{|h_{\mathrm{PT-SD}}|^2 = y} \!\!\!\!\!&=&\!\!\!\!\!  \mathrm{Pr} \left( \max_{\mathrm{SR}_i \in \mathbb{R}}\left( \min \left(\gamma_{\mathrm{SR}_i}, \gamma_{\mathrm{R}_i\mathrm{D}}\right)\right)<z \right) \nonumber \\
 &=&\!\!\!\!\!  \prod_{i=1}^N  \mathrm{Pr} \left( \min \left(\gamma_{\mathrm{SR}_i}, \gamma_{\mathrm{R}_i\mathrm{D}}\right)<z \right)  \label{eq:indep_SR_k_R_kD}\\
 &\hspace*{-1mm}=&\!\!\!\!\!\!\! \prod_{i=1}^N \!  \big[ 1 - \mathrm{Pr} \left(\gamma_{\mathrm{SR}_i} > z \right) \mathrm{Pr} \left(\gamma_{\mathrm{R}_i\mathrm{D}}>z \right) \!\big],\label{eq:12345}\nonumber \\
\end{eqnarray}\vspace*{-8mm}

\noindent where \eqref{eq:indep_SR_k_R_kD} results from the independence of $\gamma_{\mathrm{SR}_i}$ and $\gamma_{\mathrm{R}_i\mathrm{D}}$, given $y$. For ease of presentation and without compromising the insight into analysis, we assume that mean channel gains of ST-$\mathrm{SR}_i$ are the same for all relays and so is for $\mathrm{SR}_i$-SD, PT-$\mathrm{SR}_i$, and $\mathrm{SR}_i$-PD channels. Thus, we have $P_{\mathrm{SR}_i}= P_{\mathrm{SR}}$. Next, given $|h_{\mathrm{PT-SD}}|^2 = y$, we compute $\mathrm{Pr} \left(\gamma_{\mathrm{SR}_i} > z \right)$ as\vspace*{-2mm}
\begin{align}
&\mathrm{Pr} \left(\gamma_{\mathrm{SR}_i} > z \right)\nonumber \\ &= \mathrm{Pr} \left( |h_{\mathrm{ST-SR}_i}|^2 > \frac{z \left(P_{\mathrm{PT}} |h_{\mathrm{PT}-\mathrm{SR}_i}|^2 + N_0 \right)}{P_{\mathrm{ST}}}\right) \notag \\
& = \int_{0}^\infty\!\! \mathrm{Pr}\! \left(\!\! |h_{\mathrm{ST}-\mathrm{SR}_i}|^2 \!>\! \frac{z \left(P_{\mathrm{PT}}w  + N_0 \! \right)}{P_{\mathrm{ST}}}\!\right)\!\! f_{|h_{\mathrm{PT}-\mathrm{SR}_i}|^2}(w) \mathrm{d}w \notag \\ 
&= \int_{0}^\infty \exp\left( - \frac{z \left(P_{\mathrm{PT}}w  + N_0 \right)}{\Omega_{\mathrm{ST-SR}} P_{\mathrm{ST}}}\right) \frac{\exp \left(- \frac{w}{\Omega_{\mathrm{PT-SR}}} \right)}{\Omega_{\mathrm{PT-SR}}} \mathrm{d}w \nonumber \\
&= \frac{\exp\left( - \frac{z N_0 }{\Omega_{\mathrm{ST-SR}} P_{\mathrm{ST}}}\right)}{1+ \frac{z \Omega_{\mathrm{PT-SR}} P_{\mathrm{PT}}}{\Omega_{\mathrm{ST-SR}} P_{\mathrm{ST}}}}.
\label{eq:1234}
\end{align}\vspace*{-3mm}

\noindent We also compute $\mathrm{Pr} \left(\gamma_{\mathrm{R}_i\mathrm{D}}>z \right)$ as\vspace*{-4mm}

{{\small
\begin{align}
\mathrm{Pr} \left(\gamma_{\mathrm{R}_i\mathrm{D}}>z \right) &= \mathrm{Pr}\! \left(\! |h_{\mathrm{SR}_i-\mathrm{SD}}|^2 > \frac{z \left(P_{\mathrm{PT}} |h_{\mathrm{PT-SD}}|^2 + N_0 \right)}{P_{\mathrm{ST}}}\right) \nonumber \\
&= \exp\left( - \frac{z \left(P_{\mathrm{PT}}y  + N_0 \right)}{\Omega_{\mathrm{SR-SD}} P_{\mathrm{SR}}}\right).
\label{eq:123}
\end{align}}}\vspace*{-4mm}

\noindent Thus, by substituting \eqref{eq:1234} and \eqref{eq:123} in \eqref{eq:12345}, we have\vspace*{-4mm}

{{\small\begin{align}
 &F_Z(z)\big|_{|h_{\mathrm{PT-SD}}|^2 = y}\nonumber \\
 &=\Bigg(1 - \frac{\exp\left({-z N_0 \left( \frac{1}{\Omega_{\mathrm{ST-SR}} P_{\mathrm{ST}}} + \frac{1}{\Omega_{\mathrm{SR-SD}} P_{\mathrm{SR}}} \right) }\right) }{1+ \frac{z \Omega_{\mathrm{PT-SR}} P_{\mathrm{PT}}}{\Omega_{\mathrm{ST-SR}} P_{\mathrm{ST}}}}  \nonumber \\
 &\times \exp\left({- \frac{z P_{\mathrm{PT}}y }{\Omega_{\mathrm{SR-SD}} P_{\mathrm{SR}}}}\right)\Bigg)^N \notag \\ 
 &=\!\Bigg[\sum_{n=0}^N {N \choose n} (-1)^n \frac{\exp\left({-n z N_0 \left( \frac{1}{\Omega_{\mathrm{ST-SR}} P_{\mathrm{ST}}} + \frac{1}{\Omega_{\mathrm{SR-SD}} P_{\mathrm{SR}}} \right) }\right)  }{ \left( 1+ \frac{z \Omega_{\mathrm{PT-SR}} P_{\mathrm{PT}}}{\Omega_{\mathrm{ST-SR}} P_{\mathrm{ST}}} \right) ^n} \nonumber \\
&\times \exp\left({- \frac{n z P_{\mathrm{PT}}y }{\Omega_{\mathrm{SR-SD}} P_{\mathrm{SR}}}}\right)\Bigg].
\end{align}}} Hence, PDF of $Z$ is given by 
{{\small\begin{align}
 &f_Z(z)\big|_{|h_{\mathrm{PT-SD}}|^2 = y} = \sum_{n=1}^N {N \choose n} (-1)^{n+1} n \nonumber \\
 &\times \exp\left(-n z N_0 \left( \frac{1}{\Omega_{\mathrm{ST-SR}} P_{\mathrm{ST}}} + \frac{1}{\Omega_{\mathrm{SR-SD}} P_{\mathrm{SR}}} \right)\right) \nonumber \\
 &\times \exp\left(- \frac{n z P_{\mathrm{PT}}y }{\Omega_{\mathrm{SR-SD}} P_{\mathrm{SR}}}\right) \nonumber \\
 &\times \left[ \frac{ \left( \frac{N_0}{\Omega_{\mathrm{ST-SR}} P_{\mathrm{ST}}} + \frac{N_0}{\Omega_{\mathrm{SR-SD}} P_{\mathrm{SR}}} + \frac{ P_{\mathrm{PT}}y }{\Omega_{\mathrm{SR-SD}} P_{\mathrm{SR}}} \right)  }{ \left( 1+ \frac{z \Omega_{\mathrm{PT-SR}} P_{\mathrm{PT}}}{\Omega_{\mathrm{ST-SR}} P_{\mathrm{ST}}} \right) ^n} \right.\nonumber\\
 &\left. + \frac{  \frac{ \Omega_{\mathrm{PT-SR}} P_{\mathrm{PT}}}{\Omega_{\mathrm{ST-SR}} P_{\mathrm{ST}}} }{ \left( 1+ \frac{z \Omega_{\mathrm{PT-SR}} P_{\mathrm{PT}}}{\Omega_{\mathrm{ST-SR}} P_{\mathrm{ST}}} \right) ^{n+1}} \right].
\end{align}}} 
From \eqref{eq:main}, we have 
{{\small\begin{align}
&\mathrm{Pr}(\gamma_{\mathrm{tot}} < \theta_{\mathrm{S}}) \big|_{|h_{\mathrm{PT-SD}}|^2 = y} \nonumber \\
&= \int_{0}^{\theta_{\mathrm{S}}} \left( 1 - \exp\left(- \frac{\theta_{\mathrm{S}} \left(P_{\mathrm{PT}}y  + N_0 \right)}{\Omega_{\mathrm{ST-SD}} P_{\mathrm{ST}}}\right) \right.\nonumber \\
& \left.\times \exp\left(\frac{z \left(P_{\mathrm{PT}}y  + N_0 \right)}{\Omega_{\mathrm{ST-SD}} P_{\mathrm{ST}}}\right)\right)  f_Z(z)\big|_{|h_{\mathrm{PT-SD}}|^2 = y} \mathrm{d}z \notag \\
&= F_Z(\theta_{\mathrm{S}})\big|_{|h_{\mathrm{PT-SD}}|^2 = y} \nonumber\\
&- \exp\left(- \frac{\theta_{\mathrm{S}} \left(P_{\mathrm{PT}}y  + N_0 \right)}{\Omega_{\mathrm{ST-SD}} P_{\mathrm{ST}}}\right) \sum_{n=1}^N {N \choose n} (-1)^{n+1} n \nonumber\\
&\times \int_{0}^{\theta_{\mathrm{S}}} \exp\left(-z N_0 \left( \frac{n}{\Omega_{\mathrm{ST-SR}} P_{\mathrm{ST}}} + \frac{n}{\Omega_{\mathrm{SR-SD}} P_{\mathrm{SR}}} \right. \right. \nonumber \\
& \left.\left. - \frac{1}{\Omega_{\mathrm{ST-SD}} P_{\mathrm{ST}}}  \right)\right) \notag \\
&\times \frac{\exp\left(- P_{\mathrm{PT}} y z \left( \frac{n}{\Omega_{\mathrm{SR-SD}} P_{\mathrm{SR}}} - \frac{1}{\Omega_{\mathrm{ST-SD}} P_{\mathrm{ST}}}\right)\right) }{\left( 1+\frac{z \Omega_{\mathrm{PT-SR}} P_{\mathrm{PT}}}{\Omega_{\mathrm{ST-SR}} P_{\mathrm{ST}}} \right)^n}\nonumber\\
&\times \bigg( \frac{N_0}{\Omega_{\mathrm{ST-SR}} P_{\mathrm{ST}}} + \frac{N_0}{\Omega_{\mathrm{SR-SD}} P_{\mathrm{SR}}} + \frac{\frac{\Omega_{\mathrm{PT-SR}} P_{\mathrm{PT}}}{\Omega_{\mathrm{ST-SR}} P_{\mathrm{ST}} }}{\left( 1+\frac{z \Omega_{\mathrm{PT-SR}} P_{\mathrm{PT}}}{\Omega_{\mathrm{ST-SR}} P_{\mathrm{ST}}} \right) }  \nonumber \\ 
&  + \frac{P_{\mathrm{PT}} y}{\Omega_{\mathrm{SR-SD}} P_{\mathrm{SR}}} \bigg) \mathrm{d}z.
\end{align}}} Hence, the outage probability can be expressed as\vspace*{-1mm}
\begin{align}
\mathcal{P}_o = E_Y \left[ \mathrm{Pr}(\gamma_{\mathrm{tot}} < \theta_{\mathrm{S}}) \big|_{Y = y} \right] =  \mathcal{I}_1 - \mathcal{I}_2 - \mathcal{I}_3,
\end{align} where $E_Y[\cdot]$ is the expectation operator on $Y$ and\vspace*{-2mm}
\begin{align}
\mathcal{I}_1 &= \sum_{n=0}^N {N \choose n} (-1)^n \nonumber \\
& \times \frac{\exp\left(-n \theta_{\mathrm{S}} N_0 \left( \frac{1}{\Omega_{\mathrm{ST-SR}} P_{\mathrm{ST}}} + \frac{1}{\Omega_{\mathrm{SR-SD}} P_{\mathrm{SR}}} \right)\right)}{ \left( 1+ \frac{\theta_{\mathrm{S}} \Omega_{\mathrm{PT-SR}} P_{\mathrm{PT}}}{\Omega_{\mathrm{ST-SR}} P_{\mathrm{ST}}} \right) ^n}\nonumber \\
& \times \int_{0}^\infty \exp\left(- \frac{n \theta_{\mathrm{S}} P_{\mathrm{PT}}y }{\Omega_{\mathrm{SR-SD}} P_{\mathrm{SR}}}\right)\frac{\exp\left(-\frac{y}{\Omega_{\mathrm{PT-SD}}}\right)}{\Omega_{\mathrm{PT-SD}}} \mathrm{d}y,
\end{align}\vspace*{-3mm}

{{\small\begin{align}
\mathcal{I}_2 &= \exp\left(- \frac{\theta_{\mathrm{S}}  N_0 }{\Omega_{\mathrm{ST-SD}} P_{\mathrm{ST}}}\right)\sum_{n=1}^N {N \choose n} (-1)^{n+1} n \nonumber \\
& \times \int_{0}^{\theta_{\mathrm{S}}} \exp\bigg(-z N_0 \left( \frac{n}{\Omega_{\mathrm{ST-SR}} P_{\mathrm{ST}}} + \frac{n}{\Omega_{\mathrm{SR-SD}} P_{\mathrm{SR}}} \right. \nonumber \\
& \left. - \frac{1}{\Omega_{\mathrm{ST-SD}} P_{\mathrm{ST}}}  \right)\bigg) \notag \\
&\times \frac{ \left( \frac{N_0}{\Omega_{\mathrm{ST-SR}} P_{\mathrm{ST}}} + \frac{N_0}{\Omega_{\mathrm{SR-SD}} P_{\mathrm{SR}}} + \frac{\frac{\Omega_{\mathrm{PT-SR}} P_{\mathrm{PT}}}{\Omega_{\mathrm{ST-SR}} P_{\mathrm{ST}} }}{\left( 1+\frac{z \Omega_{\mathrm{PT-SR}} P_{\mathrm{PT}}}{\Omega_{\mathrm{ST-SR}} P_{\mathrm{ST}}} \right) }  \right)}{\left( 1+\frac{z \Omega_{\mathrm{PT-SR}} P_{\mathrm{PT}}}{\Omega_{\mathrm{ST-SR}} P_{\mathrm{ST}}} \right)^n} \nonumber \\
&\times\int_{y=0}^\infty \exp\bigg(- P_{\mathrm{PT}} y \left( \frac{n z}{\Omega_{\mathrm{SR-SD}} P_{\mathrm{SR}}} + \frac{\theta_{\mathrm{S}} }{\Omega_{\mathrm{ST-SD}} P_{\mathrm{ST}}} \right. \nonumber \\
& \left. - \frac{z}{\Omega_{\mathrm{ST-SD}} P_{\mathrm{ST}}}\right)\bigg)\frac{\exp\left({-\frac{y}{\Omega_{\mathrm{PT-SD}}}}\right)}{\Omega_{\mathrm{PT-SD}}} \mathrm{d}y\, \mathrm{d}z,
\end{align}}}\vspace*{-4mm}

{{\small\begin{align}
\mathcal{I}_3 &= \exp\left(- \frac{\theta_{\mathrm{S}}  N_0 }{\Omega_{\mathrm{ST-SD}} P_{\mathrm{ST}}}\right)\sum_{n=1}^N {N \choose n} (-1)^{n+1} n \nonumber \\
& \times \int_{0}^{\theta_{\mathrm{S}}} \exp\left(-z N_0 \left( \frac{n}{\Omega_{\mathrm{ST-SR}} P_{\mathrm{ST}}} + \frac{n}{\Omega_{\mathrm{SR-SD}} P_{\mathrm{SR}}} \right. \right. \nonumber \\
& \left. \left. - \frac{1}{\Omega_{\mathrm{ST-SD}} P_{\mathrm{ST}}}  \right)\right)\frac{ \frac{P_{\mathrm{PT}}}{\Omega_{\mathrm{SR-SD}} P_{\mathrm{SR}}}}{\left( 1+\frac{z \Omega_{\mathrm{PT-SR}} P_{\mathrm{PT}}}{\Omega_{\mathrm{ST-SR}} P_{\mathrm{ST}}} \right)^n} \notag \\
& \times  \int_{0}^\infty \!\! y \exp\left(\!- P_{\mathrm{PT}} y \left( \frac{n z}{\Omega_{\mathrm{SR-SD}} P_{\mathrm{SR}}} + \frac{\theta_{\mathrm{S}} }{\Omega_{\mathrm{ST-SD}} P_{\mathrm{ST}}} \right. \right. \nonumber \\
& \left. \left.- \frac{z}{\Omega_{\mathrm{ST-SD}} P_{\mathrm{ST}}}\right)\right) \frac{\exp\left(-\frac{y}{\Omega_{\mathrm{PT-SD}}}\right)}{\Omega_{\mathrm{PT-SD}}} \mathrm{d}y\, \mathrm{d}z.
\end{align}}} 
We use the following results in \eqref{eq:expec_exp_y} and \eqref{eq:expec_y_exp_y} to derive the integrations $\mathcal{I}_i, i=1,2,3$: When $Y$ is an exponential random variable with mean $\Omega_Y$, we have
\begin{align}
E_Y \left[ \exp\left({-RY}\right)\right] &=\frac{1}{\Omega_Y} \int_{0}^\infty \exp\left({- \left( R + \frac{1}{\Omega_Y} \right)}\right) \mathrm{d}y \nonumber \\
&= \frac{1}{1+ \Omega_Y R},  \label{eq:expec_exp_y}
\end{align}
\begin{align}
E_Y \left[ Y \exp\left({-RY}\right)\right] &=\frac{1}{\Omega_Y} \int_{0}^\infty y \exp\left({- \left( R + \frac{1}{\Omega_Y} \right)}\!\right) \mathrm{d}y \nonumber \\
&= \frac{\Omega_Y}{ \left( 1+ \Omega_Y R \right)^2}, \label{eq:expec_y_exp_y}
\end{align} 
with $R \geq 0$. Using \eqref{eq:expec_exp_y}, we compute $\mathcal{I}_1$ as\vspace*{-1mm}
\begin{align}
\mathcal{I}_1 &= \sum_{n=0}^N {N \choose n} (-1)^n \nonumber \\
& \times \frac{\exp\left({-n \theta_{\mathrm{S}} N_0 \left( \frac{1}{\Omega_{\mathrm{ST-SR}} P_{\mathrm{ST}}} + \frac{1}{\Omega_{\mathrm{SR-SD}} P_{\mathrm{SR}}} \right)}\right)}{ \left( 1+ \frac{\theta_{\mathrm{S}} \Omega_{\mathrm{PT-SR}} P_{\mathrm{PT}}}{\Omega_{\mathrm{ST-SR}} P_{\mathrm{ST}}} \right)^n \left( 1+ \frac{n \theta_{\mathrm{S}} \Omega_{\mathrm{PT-SD}} P_{\mathrm{PT}}}{\Omega_{\mathrm{SR-SD}} P_{\mathrm{SR}}} \right)}.
\end{align} To compute $\mathcal{I}_2$, we write it as
\begin{align}
\mathcal{I}_2 &= \exp\left({- \frac{\theta_{\mathrm{S}}  N_0 }{\Omega_{\mathrm{ST-SD}} P_{\mathrm{ST}}}}\right)\sum_{n=1}^N n {N \choose n} (-1)^{n+1} \nonumber \\
& \times \left( \mathcal{I}_{2,1,n} + \mathcal{I}_{2,2,n} \right),
\end{align} where\vspace*{-1mm}
\begin{eqnarray}
\mathcal{I}_{2,1,n} &=& \int_{0}^{\theta_{\mathrm{S}}} \frac{\left( \frac{N_0}{\Omega_{\mathrm{ST-SR}} P_{\mathrm{ST}}} + \frac{N_0}{\Omega_{\mathrm{SR-SD}} P_{\mathrm{SR}}} \right) }{\left( 1+\frac{z \Omega_{\mathrm{PT-SR}} P_{\mathrm{PT}}}{\Omega_{\mathrm{ST-SR}} P_{\mathrm{ST}}} \right)^n }  \nonumber \\
&&\!\hspace*{-19mm} \times \frac{ \exp\!\left(\!-z N_0 \left( \frac{n}{\Omega_{\mathrm{ST-SR}} P_{\mathrm{ST}}} + \frac{n}{\Omega_{\mathrm{SR-SD}} P_{\mathrm{SR}}} - \frac{1}{\Omega_{\mathrm{ST-SD}} P_{\mathrm{ST}}}  \right)\right)\mathrm{d}z}{\left(\! 1 + \Omega_{\mathrm{PT-SD}} P_{\mathrm{PT}}\!\left(\! \frac{n z}{\Omega_{\mathrm{SR-SD}} P_{\mathrm{SR}}} \!+\!\! \frac{\theta_{\mathrm{S}} }{\Omega_{\mathrm{ST-SD}} P_{\mathrm{ST}}} - \frac{z}{\Omega_{\mathrm{ST-SD}} P_{\mathrm{ST}}}\right) \!\right)} \nonumber \\
\end{eqnarray}\vspace*{-6mm}

\noindent and\vspace*{-1mm}
\begin{eqnarray}
\mathcal{I}_{2,2,n} &=& \int_{0}^{\theta_{\mathrm{S}}} \frac{\frac{\Omega_{\mathrm{PT-SR}} P_{\mathrm{PT}}}{\Omega_{\mathrm{ST-SR}} P_{\mathrm{ST}}}   }{\left( 1+\frac{z \Omega_{\mathrm{PT-SR}} P_{\mathrm{PT}}}{\Omega_{\mathrm{ST-SR}} P_{\mathrm{ST}}} \right)^{n+1} } \nonumber \\
&&\!\!\hspace*{-19mm} \times \frac{ \exp\!\left(\!-z N_0\! \left(\! \frac{n}{\Omega_{\mathrm{ST-SR}} P_{\mathrm{ST}}} + \frac{n}{\Omega_{\mathrm{SR-SD}} P_{\mathrm{SR}}} - \frac{1}{\Omega_{\mathrm{ST-SD}} P_{\mathrm{ST}}}  \right)\right)\mathrm{d}z}{\left(\! 1 + \Omega_{\mathrm{PT-SD}} P_{\mathrm{PT}}\!\left(\! \frac{n z}{\Omega_{\mathrm{SR-SD}} P_{\mathrm{SR}}} \!+\!\! \frac{\theta_{\mathrm{S}} }{\Omega_{\mathrm{ST-SD}} P_{\mathrm{ST}}} - \frac{z}{\Omega_{\mathrm{ST-SD}} P_{\mathrm{ST}}}\right) \!\!\right)}. \nonumber \\
\end{eqnarray}\vspace*{-7mm}

\noindent To compute $\mathcal{I}_{2,1,n}$ and $\mathcal{I}_{2,2,n}$, we use the following notations for convenience of presentation:\vspace*{-1mm}
\begin{align}
S &= N_0 \left(\! \frac{n}{\Omega_{\mathrm{ST-SR}} P_{\mathrm{ST}}} + \frac{n}{\Omega_{\mathrm{SR-SD}} P_{\mathrm{SR}}} - \frac{1}{\Omega_{\mathrm{ST-SD}} P_{\mathrm{ST}}}  \!\right), \notag \\
\mu &= \Omega_{\mathrm{PT-SD}} P_{\mathrm{PT}} \left( \frac{n}{\Omega_{\mathrm{SR-SD}} P_{\mathrm{SR}}} - \frac{1}{\Omega_{\mathrm{ST-SD}} P_{\mathrm{ST}}} \right), \notag \\
\tau &= \frac{\frac{\Omega_{\mathrm{PT-SD}} P_{\mathrm{PT}} \theta_{\mathrm{S}}}{\Omega_{\mathrm{ST-SD}} P_{\mathrm{ST}}} +1 }{\mu}, \nonumber \\
 \pi_1 &= \left( \frac{\Omega_{\mathrm{ST-SR}} P_{\mathrm{ST}}}{ \Omega_{\mathrm{PT-SR}} P_{\mathrm{PT}}} \right).
\end{align} 
Thus, we can write\vspace*{-1mm}
\begin{align}
\mathcal{I}_{2,1,n} &= \frac{\pi_1^n}{\mu} \left( \frac{N_0}{\Omega_{\mathrm{ST-SR}} P_{\mathrm{ST}}} + \frac{N_0}{\Omega_{\mathrm{SR-SD}} P_{\mathrm{SR}}} \right) \nonumber \\
& \times \underbrace{\int_{0}^{\theta_{\mathrm{S}}} \frac{\exp({-Sz})}{ \left( z + \pi_1 \right)^n (z + \tau)}}_{\mathcal{J}_{2,1,n}} \mathrm{d}z.
\label{eq:11}
\end{align}\vspace*{-3mm}

\noindent For $\Omega_{\mathrm{ST-SD}} > \Omega_{\mathrm{ST-SR}}$ and $\Omega_{\mathrm{ST-SD}} > \Omega_{\mathrm{SR-SD}}$, we have $S>0, \mu>0, \tau>0$ and we can write \eqref{eq:11} in terms of the exponential integral as shown later in this section. Using the substitution, $r=z + \pi_1$ and denoting $\chi = \tau - \pi_1$, we write\vspace*{-4mm}

\begin{align}
\mathcal{J}_{2,1,n} = \exp({S\pi_1}) \int_{\pi_1}^{\pi_1+\theta_{\mathrm{S}}} \frac{\exp({-Sr})}{ r^n (r+ \chi)} \mathrm{d}r.
\end{align} Using the partial fraction expansion, we have
\begin{equation}
\frac{1}{r^n (r+\chi)} = \sum_{m=0}^{n-1} \frac{(-1)^m }{ \chi^{m+1} r^{n-m}} + \frac{1}{(-\chi)^n (r+ \chi)},
\end{equation} Thus, we can write
{{\small\begin{align}
\mathcal{J}_{2,1,n} &= \exp({S\pi_1}) \!\!\int_{\pi_1}^{\pi_1+\theta_{\mathrm{S}}}\! \exp({-Sr}) \nonumber\\
&\times \left( \sum_{m=0}^{n-1} \frac{(-1)^m}{ \chi^{m+1} r^{n-m}} + \frac{1}{(-\chi)^n (r+ \chi)} \right) \mathrm{d}r \notag \\ 
&= \exp({S\pi_1}) \sum_{m=0}^{n-1} \frac{(-1)^m}{ \chi^{m+1}} \int_{\pi_1}^{\pi_1+\theta_{\mathrm{S}}} \frac{ \exp({-Sr})}{r^{n-m}} \mathrm{d}r \nonumber \\
&+ \frac{\exp({S\pi_1}) }{(-\chi)^n} \int_{\pi_1+ \chi}^{\pi_1+\chi+\theta_{\mathrm{S}}} \frac{ \exp({-S(p-\chi)})}{p} \mathrm{d}p \label{j21n_sub1}\\
&=\exp({S\pi_1}) \sum_{m=0}^{n-1} \frac{(-1)^m}{ \chi^{m+1}} S^{n-m-1} \int_{S \pi_1}^{S(\pi_1+\theta_{\mathrm{S}})} \frac{ \exp({-z})}{z^{n-m}} \mathrm{d}z \nonumber \\
&+ \frac{\exp({S(\pi_1 + \chi)}) }{(-\chi)^n} \int_{S \tau}^{S(\tau+\theta_{\mathrm{S}})} \frac{ \exp({-y})}{y} \mathrm{d}y \label{j21n_sub2}\\
&= \exp({S\pi_1}) \bigg(\sum_{m=0}^{n-2} \frac{(-1)^m}{ \chi^{m+1}} S^{n-m-1} \nonumber \\
&\times \left[ \Gamma(m-n+1, S \pi_1) - \Gamma(m-n+1, S (\pi_1+\theta_{\mathrm{S}})) \right]\bigg) \notag \\
&+ \exp({S\pi_1}) \frac{(-1)^{n-1}}{ \chi^{n}} \left[ \mathrm{E_1}(S\pi_1)- \mathrm{E_1}(S (\pi_1+\theta_{\mathrm{S}})) \right]\nonumber\\
& + \exp({S\tau}) \frac{(-1)^{n}}{ \chi^{n}} \left[ \mathrm{E_1}(S\tau)- \mathrm{E_1}(S (\tau+\theta_{\mathrm{S}})) \right], 
\end{align}}}\vspace*{-4mm}

\noindent where in \eqref{j21n_sub1}, we use the substitution $p = r+ \chi$, and in \eqref{j21n_sub2}, we use $z=Sr$ and $y=Sp$; $\Gamma( . , .)$ and $\mathrm{E_1}(\cdot)$ are upper incomplete gamma function and exponential integral \cite{gradshteyn}, respectively with $\mathrm{E_1}(x) = \int_{x}^{\infty}\frac{\exp(-t)}{t}\mathrm{d}t$. Similarly, we compute $\mathcal{I}_{2,2,n}$ by representing it as\vspace*{-1mm}
\begin{equation}
\mathcal{I}_{2,2,n} = \frac{\pi_1^n}{\mu} \underbrace{\int_{0}^{\theta_{\mathrm{S}}} \frac{\exp({-Sz})}{ \left( z + \pi_1 \right)^{n+1} (z + \tau)}}_{\mathcal{J}_{2,2,n}} \mathrm{d}z,
\end{equation} where $\mathcal{J}_{2,2,n}$ is computed as
\begin{align}
\mathcal{J}_{2,2,n} &= \exp({S\pi_1})\bigg( \sum_{m=0}^{n-1} \frac{(-1)^m}{ \chi^{m+1}} S^{n-m}\nonumber \\
& \times \left[ \Gamma(m-n, S \pi_1) - \Gamma(m-n, S (\pi_1+\theta_{\mathrm{S}})) \right]\bigg) \notag \\
& + \exp({S\pi_1}) \frac{(-1)^{n}}{ \chi^{n+1}} \left[ \mathrm{E_1}(S\pi_1)- \mathrm{E_1}(S (\pi_1+\theta_{\mathrm{S}})) \right] \nonumber \\
& + \exp({S\tau}) \frac{(-1)^{n+1}}{ \chi^{n+1}} \left[ \mathrm{E_1}(S\tau)- \mathrm{E_1}(S (\tau+\theta_{\mathrm{S}})) \right].
\end{align} 
We note that $\mathcal{J}_{2,2,n} = \mathcal{J}_{2,1,n}$ with $n$ replaced by $n+1$. Thus, we can use the same procedure to compute both these expressions.
Using \eqref{eq:expec_y_exp_y}, we write $\mathcal{I}_3$ as\vspace*{-3mm}

{{\small\begin{eqnarray}
\mathcal{I}_3 &=& \exp\left({- \frac{\theta_{\mathrm{S}}  N_0 }{\Omega_{\mathrm{ST-SD}} P_{\mathrm{ST}}}}\right)\sum_{n=1}^N {N \choose n} (-1)^{n+1} n \nonumber \\
&&\hspace*{-14mm}\times \int_{0}^{\theta_{\mathrm{S}}} \!\!\!\!\exp\!\left(\!\!{-z N_0\! \left( \frac{n}{\Omega_{\mathrm{ST-SR}} P_{\mathrm{ST}}} + \frac{n}{\Omega_{\mathrm{SR-SD}} P_{\mathrm{SR}}} - \frac{1}{\Omega_{\mathrm{ST-SD}} P_{\mathrm{ST}}}  \right)}\!\!\right) \notag \\
&&\hspace*{-14mm}\frac{ \frac{P_{\mathrm{PT}}}{\Omega_{\mathrm{SR-SD}} P_{\mathrm{SR}}}}{\left( 1+\frac{z \Omega_{\mathrm{PT-SR}} P_{\mathrm{PT}}}{\Omega_{\mathrm{ST-SR}} P_{\mathrm{ST}}} \right)^n} \nonumber\\
&&\hspace*{-14mm}\times\frac{\Omega_{\mathrm{PT-SD}}\,\mathrm{d}z }{\left(\! 1 + \Omega_{\mathrm{PT-SD}} P_{\mathrm{PT}}\! \left( \!\frac{n z}{\Omega_{\mathrm{SR-SD}} P_{\mathrm{SR}}} + \frac{\theta_{\mathrm{S}} }{\Omega_{\mathrm{ST-SD}} P_{\mathrm{ST}}} - \frac{z}{\Omega_{\mathrm{ST-SD}} P_{\mathrm{ST}}}\right)\! \right)^2}\notag \\
&\hspace*{-14mm}=&\hspace*{-9mm}\frac{\Omega_{\mathrm{PT-SD}} P_{\mathrm{PT}}}{\Omega_{\mathrm{SR-SD}} P_{\mathrm{SR}}} \exp\left({- \frac{\theta_{\mathrm{S}}  N_0 }{\Omega_{\mathrm{ST-SD}} P_{\mathrm{ST}}}}\right)\sum_{n=1}^N {N \choose n} (-1)^{n+1} n \frac{\pi_1^n}{\mu^2} \nonumber \\
&&\hspace*{-13mm} \times \int_{0}^{\theta_{\mathrm{S}}} \underbrace{\frac{\exp({-Sz})}{ \left( z + \pi_1 \right)^{n} (z + \tau)^2}}_{\mathcal{J}_{3,n}} \mathrm{d}z,
\end{eqnarray}}}\vspace*{-4mm}

\noindent where we use the same notations as for the case of $\mathcal{I}_2$. With the substitution of $t=z+\pi_1$, $\mathcal{J}_{3,n}$ can be written as
\begin{equation}
\mathcal{J}_{3,n} = \exp({S \pi_1}) \int_{\pi_1}^{\pi_1+\theta_{\mathrm{S}}} \frac{\exp({-St})}{ t^{n} (t + \chi)^2} \mathrm{d}t
\end{equation} For computation of $\mathcal{J}_{3,n}$, we use the following partial fraction expansion:
\begin{eqnarray}
\frac{1}{r^n (r+\chi)^2} &=& \left(\sum_{m=0}^{n-1} \frac{(-1)^m (m+1)}{\chi^{m+2} r^{n-m}}\right) +  \frac{1}{(-\chi)^n (r+ \chi)^2} \nonumber\\
&& - \frac{n}{(-\chi)^{n+1} (r+ \chi)}.
\end{eqnarray}
Using the steps similar to that of the derivation of $\mathcal{I}_{2,1,n}$ given in \eqref{eq:11}, we, hereby, can write the expression of $\mathcal{J}_{3,n}$ as
\begin{eqnarray}
\mathcal{J}_{3,n} &\!=& \exp({S\pi_1}) \Bigg(\sum_{m=0}^{n-2} \frac{(-1)^m (m+1)}{ \chi^{m+2}} S^{n-m-1} \nonumber\\
&&\hspace*{-7mm} \times \left[ \Gamma(m-n+1, S \pi_1) - \Gamma(m-n+1, S (\pi_1+\theta_{\mathrm{S}})) \right]\Bigg) \notag \\
&& \hspace*{-7mm}+ \exp({S\pi_1}) \frac{(-1)^{n-1}}{ \chi^{n+1}} \left[ \mathrm{E_1}(S\pi_1)- \mathrm{E_1}(S (\pi_1+\theta_{\mathrm{S}})) \right] \notag \\ 
&& \hspace*{-7mm}+ \exp({S\tau}) \frac{(-1)^{n}}{ \chi^{n}} \left[ \Gamma(-1, S \tau) - \Gamma(-1, S (\tau+\theta_{\mathrm{S}})) \right] \notag \\ 
&&\hspace*{-7mm}+ \exp({S\tau}) \frac{(-1)^{n} n}{ \chi^{n+1}} \left[ \mathrm{E_1}(S\tau)- \mathrm{E_1}(S (\tau+\theta_{\mathrm{S}})) \right].
\end{eqnarray}
\section{Results and Discussions}
Using the analysis performed in previous sections, we investigate the effects of direct link, primary interference, primary outage constraint, and the peak power constraint on the outage performance of the secondary system. We also validate the analysis by simulation results. The simulation parameters are as follow: $\Omega_{\mathrm{ST-SD}} = \mathrm{1.5}$, $\Omega_{\mathrm{PT-PD}} = \Omega_{\mathrm{ST-SR}} = \Omega_{\mathrm{SR-SD}} = \mathrm{1}$; $\Omega_{\mathrm{PT-SR}} = \Omega_{\mathrm{PT-SD}} = \Omega_{\mathrm{ST-PD}} = \Omega_{\mathrm{SR-PD}} = \mathrm{0.5}$, $N_0 = \mathrm{1}$, $R_{\mathrm{p}} = \mathrm{0.4}\mathrm{bits/s/Hz}$, $R_{\mathrm{S}} = \mathrm{0.1}\mathrm{bits/s/Hz}$. 

\begin{figure}
\centering
\includegraphics[scale=0.34]{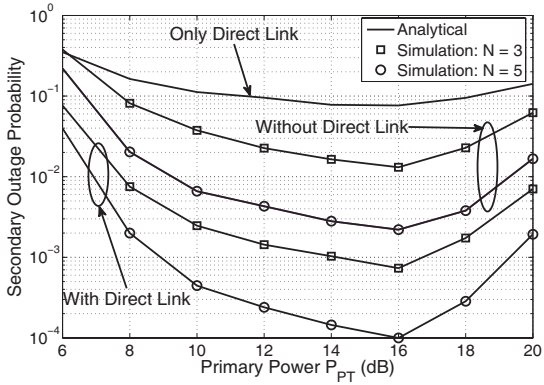}\vspace*{-3mm}
\caption{Secondary outage probability vs. Primary power ($P_{\mathrm{PT}}$) for different number of relays $N$, with and without direct link, $P_{\mathrm{pk}} = 15\mathrm{dB}$, $\lambda_{\mathrm{p}} = 10^{-1}$.}
\label{fig:22}\vspace*{-4mm}
\end{figure}

Fig.\,\ref{fig:22} shows the effect of primary power $P_{\mathrm{PT}}$ on the secondary outage probability $\mathcal{P}_{o}$. The increase in $P_{\mathrm{PT}}$ has two opposite effects on $\mathcal{P}_{o}$: 1) It improves the quality of the primary link, in turn, increases SINR at the primary destination. This leads to decrease in the primary outage probability, providing an extra margin for transmit powers of secondary transmitter ST ($P_{\mathrm{ST}}$) and the selected relay SR ($P_{\mathrm{SR}}$), which further helps in reducing the secondary outage probability; 2) it increases the interference to the secondary system, thereby increasing the secondary outage probability. From Fig.\,\ref{fig:22}, we can observe that, initially, the secondary outage probability reduces as $P_{\mathrm{PT}}$ increases. However, if $P_{\mathrm{PT}}$ is increased beyond a level, the peak power constraint is reached for SU, which does not allow further increase in $P_{\mathrm{ST}}$ and $P_{\mathrm{SR}}$. Thus, with an additional increase in the primary power, SINR at the secondary destination reduces as $P_{\mathrm{ST}}$ and $P_{\mathrm{SR}}$ cannot be increased further, degrading SU's outage performance. We can also see from Fig.\,\ref{fig:22} that the presence of direct link effectively helps in improving SU's performance. Also, the increase in the number of relays improves secondary's outage performance due to the increase in the diversity gain.

\begin{figure}
\centering
\includegraphics[scale=0.34]{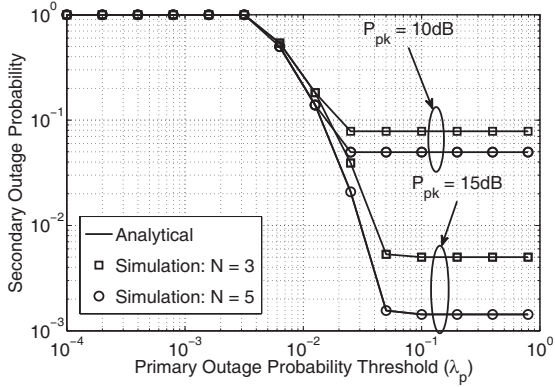}\vspace*{-3mm}
\caption{Secondary outage probability vs. Primary outage probability threshold ($\lambda_{\mathrm{p}}$) for different values of peak power constraint $P_{\mathrm{pk}}$ and number of relays $N$, $P_{\mathrm{PT}} = 20\mathrm{dB}$.}
\label{fig:33}\vspace*{-7mm}
\end{figure}

Fig.\,\ref{fig:33} shows the effect of the primary outage probability threshold $\lambda_{\mathrm{p}}$ on the secondary outage probability. We can see that increase in $\lambda_{\mathrm{p}}$ relaxes the constraint on $P_{\mathrm{ST}}$ and $P_{\mathrm{SR}}$. But, if we increase $\lambda_{\mathrm{p}}$ beyond a level, the peak power constraint is reached, and ST and SR may transmit with the maximum power $P_{\mathrm{pk}}$ even though they are allowed, by the primary, to transmit with higher power than $P_{\mathrm{pk}}$. In this case, unlike in Fig.\,\ref{fig:22}, the primary power, in turn, the primary interference to SD remains constant. Thus, irrespective of the increase in $\lambda_{\mathrm{p}}$, the secondary outage probability remains constant$-$we call it as floor$-$once the peak power constraint is reached. We can also notice from Fig.\,\ref{fig:33} that relaxing the peak power constraint delays the arrival of the floor as expected.\vspace*{-1mm}

\bibliographystyle{ieeetr}
\bibliography{paper}

\begin{thebibliography}{10}

\bibitem{mitola}
J.~Mitola and G.~Q. Maguire, ``Cognitive radio: Making software more
  personal,'' {\em IEEE Pers. Commun.}, vol.~6, pp.~13--18, Aug. 1999.

\bibitem{qing}
Q.~Zhao and B.~Sadler, ``A survey of dynamic spectrum access,'' {\em IEEE
  Signal Process. Mag.}, vol.~24, pp.~79--89, May 2007.

\bibitem{xing}
Y.~Xing, C.~Mathur, M.~Haleem, R.~Chandramouli, and K.~Subbalakshmi, ``Dynamic
  spectrum access with {QoS} and interference temperature constraints,'' {\em
  IEEE Trans. Mobile Comput.}, vol.~6, pp.~423--433, Apr. 2007.

\bibitem{luo}
L.~Luo, P.~Zhang, G.~Zhang, and J.~Qin, ``Outage performance for cognitive
  relay networks with underlay spectrum sharing,'' {\em IEEE Commun. Lett.},
  vol.~15, pp.~710--712, July 2011.

\bibitem{zhang1}
Q.~Zhang, J.~Jia, and J.~Zhang, ``Cooperative relay to improve diversity in
  cognitive radio networks,'' {\em IEEE Commun. Mag.}, vol.~47, pp.~111--117,
  Feb. 2009.

\bibitem{zou}
Y.~Zou, J.~Zhu, B.~Zheng, and Y.-D. Yao, ``An adaptive cooperation diversity
  scheme with best-relay selection in cognitive radio networks,'' {\em IEEE
  Trans. Signal Process.}, vol.~58, pp.~5438--5445, Oct. 2010.

\bibitem{duong}
T.~Duong, V.~Bao, and H.-J. Zepernick, ``Exact outage probability of cognitive
  {AF} relaying with underlay spectrum sharing,'' {\em Electron. Lett.},
  vol.~47, pp.~1001--1002, Aug. 2011.

\bibitem{lee}
J.~Lee, H.~Wang, J.~Andrews, and D.~Hong, ``Outage probability of cognitive
  relay networks with interference constraints,'' {\em IEEE Trans. Wireless
  Commun.}, vol.~10, pp.~390--395, Feb. 2011.

\bibitem{sun}
C.~Sun and K.~Letaief, ``User cooperation in heterogeneous cognitive radio
  networks with interference reduction,'' in {\em Proc. IEEE ICC},
  pp.~3193--3197, May 2008.

\bibitem{li}
L.~Li, X.~Zhou, H.~Xu, G.~Li, D.~Wang, and A.~Soong, ``Simplified relay
  selection and power allocation in cooperative cognitive radio systems,'' {\em
  IEEE Trans. Wireless Commun.}, vol.~10, pp.~33--36, Jan. 2011.

\bibitem{khaled}
K.~Letaief and W.~Zhang, ``Cooperative communications for cognitive radio
  networks,'' {\em Proc. IEEE}, vol.~97, pp.~878--893, May 2009.

\bibitem{si11}
J.~Si, Z.~Li, X.~Chen, B.~Hao, and Z.~Liu, ``On the performance of cognitive
  relay networks under primary user's outage constraint,'' {\em IEEE Commun.
  Lett.}, vol.~15, pp.~422--424, April 2011.

\bibitem{duong2}
T.~Duong, V.~Bao, G.~Alexandropoulos, and H.-J. Zepernick, ``Cooperative
  spectrum sharing networks with {AF} relay and selection diversity,'' {\em
  Electron. Lett.}, vol.~47, pp.~1149--1151, Sept. 2011.

\bibitem{duong3}
T.~Duong, V.~Bao, H.~Tran, G.~Alexandropoulos, and H.-J. Zepernick, ``Effect of
  primary network on performance of spectrum sharing {AF} relaying,'' {\em
  Electron. Lett.}, vol.~48, pp.~25--27, Jan. 2012.

\bibitem{xu}
W.~Xu, J.~Zhang, P.~Zhang, and C.~Tellambura, ``Outage probability of
  decode-and-forward cognitive relay in presence of primary user's
  interference,'' {\em IEEE Commun. Lett.}, vol.~16, pp.~1252--1255, Aug. 2012.

\bibitem{yang}
P.~Yang, L.~Luo, and J.~Qin, ``Outage performance of cognitive relay networks
  with interference from primary user,'' {\em IEEE Commun. Lett.}, vol.~16,
  pp.~1695--1698, Oct. 2012.

\bibitem{huang}
H.~Huang, Z.~Li, J.~Si, and R.~Gao, ``Outage analysis of underlay cognitive
  multiple relays networks with a direct link,'' {\em IEEE Commun. Lett.},
  vol.~17, pp.~1600--1603, Aug. 2013.

\bibitem{yan}
Z.~Yan, X.~Zhang, and W.~Wang, ``Exact outage performance of cognitive relay
  networks with maximum transmit power limits,'' {\em IEEE Commun. Lett.},
  vol.~15, pp.~1317--1319, Dec. 2011.

\bibitem{shah}
V.~Shah, N.~Mehta, and R.~Yim, ``The relay selection and transmission trade-off
  in cooperative communication systems,'' {\em IEEE Trans. Wireless Commun.},
  vol.~9, pp.~2505--2515, Aug. 2010.

\bibitem{ikki}
S.~Ikki and M.~Ahmed, ``Performance analysis of cooperative diversity wireless
  networks over {Nakagami}-m fading channel,'' {\em IEEE Commun. Lett.},
  vol.~11, pp.~334--336, April 2007.

\bibitem{si}
J.~Si, Z.~Li, J.~Chen, P.~Qi, and H.~Huang, ``Performance analysis of adaptive
  modulation in cognitive relay networks with interference constraints,'' in
  {\em Proc. IEEE WCNC}, pp.~2631--2636, April 2012.

\bibitem{gradshteyn}
I.~Gradshteyn and I.~Ryzhik, {\em Table of Integrals, Series and Products}.
\newblock Academic Press, {7th}~ed., 2007.

\end{thebibliography}
\end{document}